\documentclass[%
 twocolumn,
 reprint,
showpacs,
 amsmath,amssymb,
 aps,
]{revtex4}
\usepackage{color}
\usepackage{graphicx}
\usepackage{dcolumn}
\usepackage{bm}

\begin{document}

\preprint{APS/123-QED\author{Takahiro Ohgoe}}

\title{Quantum phases of hard-core bosons on 2D lattices with anisotropic dipole-dipole interaction}

\author{Takahiro Ohgoe$^1$}
\author{Takafumi Suzuki$^2$}
\author{Naoki Kawashima$^1$}%
\affiliation{%
	$^1$Institute for Solid State Physics, University of Tokyo, Kashiwa, Chiba 277-8581, Japan\\
	$^2$Research Center for Nano-Micro Structure Science and Engineering, Graduate School of Engineering, University of Hyogo, Himeji, Hyogo 671-2280, Japan}%

\date{\today}

\begin{abstract}
By using an unbiased quantum Monte Carlo method, we investigate the hard-core Bose-Hubbard model on a square lattice with anisotropic dipole-dipole interaction. To study the effect of the anisotropy, dipole moments are assumed to be polarized in the $y$ direction on the two-dimensional (2D) $xy$ plane. To perform efficient simulations of long-range interacting systems, we use the worm algorithm with an $O(N)$ Monte Carlo method. We obtain the ground-state phase diagram that includes a superfluid phase and a striped solid phase at half-filling as two main phases. In addition to these two main phases, we find a small region where there are multi plateaus in the particle density for small hopping amplitudes. In this region, the number of plateaus increases as the system size increases. This indicates the appearance of numerous competing ground states due to frustrated interactions. In our simulations, we find no evidence of a supersolid phase.

\end{abstract}

\maketitle


\section{\label{sec:level1}Introduction}
	Since the experimental realization of a Bose-Einstein condense (BEC) of $^{52}$Cr with a large permanent magnetic dipole moment\cite{griesmaier2005}, systems with the dipole-dipole interaction have attracted great interact. This is because long-range and anisotropic nature of the  dipole-dipole interaction shows fascinating phenomena that are different from those of short-range and isotropic interactions. The observation of $d$-wave collapse of a $^{52}$Cr BEC is an interesting example\cite{lahaye2008}. More recently, a BEC of $^{168}$Er with a larger magnetic dipole moment has also been realized\cite{aikawa2012}. Furthermore, there are great experimental efforts toward the realization of a system of polar molecules with field-induced electric dipole moments\cite{sage2005, ni2008, ospelkaus2008}. 

	In previous theoretical and numerical works, novel quantum phases of dipolar bosons such as supersolid phases have been predicted in optical lattice systems\cite{otterlo1995, batrouni2000, goral2002, wessel2005, boninsegni2005, sengupta2005, batrouni2006, yi2007, suzuki2007, yamamoto2009, danshita2009, capogrosso2010-1, pollet2010, xi2011, ohgoe2011_2, ohgoe2012, yamamoto2012, ohgoe2012_2}. In particular, by recent quantum Monte Carlo simulations, supersoild phases have been found in the hard-core bosons on a square lattice\cite{capogrosso2010-1} and a triangular lattice\cite{pollet2010} with the dipole-dipole interaction. The Hamiltonian considered is given by
	\begin{eqnarray}
		H & = & - t \sum_{\langle i, j\rangle} ( b_{i}^{\dagger} b_{j} + h.c. ) - \mu \sum_{i} n_{i} + \sum_{i<j} V_{ij} n_i n_j. \label{eq:hamiltonian}
	\end{eqnarray}
	Here, $b^{\dagger}_{i}$($b_{i}$) is the bosonic creation (annihilation) operator on a site $i$, and $n_{i}$ is the particle number operator defined by $n_{i} = b^{\dagger}_{i} b_{i}$. The first, second and third term describe the kinetic energy with hopping amplitude $t$, the chemical potential, and the dipole-dipole interaction, respectively. More specifically, the dipole-dipole interaction $V_{ij}$ is given by
	\begin{eqnarray}
		V_{ij} & = & V \frac{r_{ij}^2 (\mbox{\boldmath $e$}_{i} \cdot \mbox{\boldmath $e$}_{j}) - 3(\mbox{\boldmath $e$}_{i} \cdot \mbox{\boldmath $r$}_{ij})(\mbox{\boldmath $e$}_{j} \cdot \mbox{\boldmath $r$}_{ij}) }{r_{ij}^5}, \label{eq:dipole}
	\end{eqnarray}
	where $\mbox{\boldmath $r$}_{ij}=\mbox{\boldmath $r$}_{i}-\mbox{\boldmath $r$}_{j}$ is the relative position vector between sites $i$ and $j$, and $\mbox{\boldmath $e$}_{i}$ is a unit vector of dipole moment on a site $i$. The strength $V$ of the dipole-dipole interaction is given by $V=\mu_{0} \mu_{m}^2/4\pi$ for a magnetic dipole moment $\bm{\mu}_{m}=\mu_{m} \mbox{\boldmath $e$}$ and $V=d^{2}/4\pi \epsilon_{0}$ for an electric dipole moment $\mbox{\boldmath $d$}=d \mbox{\boldmath $e$}$. Here, $\mu_{0}$ and $\epsilon_{0}$ is the permeability and permittivity of vacuum, respectively. The authors of Refs. \cite{capogrosso2010-1} and \cite{pollet2010} have studied the case where dipole moments are uniformly perpendicular to the 2D plane, {\it i.e.} $\mbox{\boldmath $e$}=(0,0,1)$. In such a case, the dipole-dipole interaction reduces to the purely repulsive one $V_{ij}=V/r^{3}$. Thus, the anisotropic nature of the dipole-dipole interaction is absent. When the dipole moments are tilted, the dipole-dipole interaction shows anisotropy with attractive interactions as well as repulsive ones. An interesting question is how the anisotropy changes the phase diagram and whether a supersolid phase is also found.

\begin{figure}[h]
		\includegraphics[width=7cm]{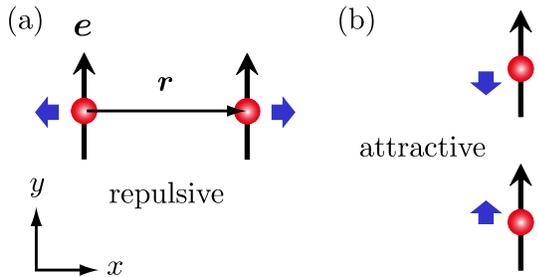}
		\caption{\label{fig:ddi} (Color online) (a) and (b) Dipole-dipole interactions in two different situations where the relative position vector $\mbox{\boldmath $r$}$ between two particles are perpendicular and parallel to the dipole moments, respectively. The dipole-dipole interaction is repulsive ($V/r^{3}$) in (a), but it becomes attractive ($-2V/r^{3}$) in (b).}
	\end{figure}

	In this paper, we therefore simply consider the case where dipole moments are polarized in the $y$ direction on the 2D $xy$ plane by an external uniform field in order to study anisotropic properties of the model.. Since the unit vector of dipole moments is given by $\mbox{\boldmath $e$}=(0,1,0)$ in this situation, the dipole-dipole interaction reduces to 
	\begin{eqnarray}
		V_{ij} & = & \frac{V}{r^3} \left( 1 - \frac{3r_{y}^2}{r^2} \right),
	\end{eqnarray}
	where $r$ is an abbreviation of $r_{ij}$ and $r_{y}$ is the distance between two particles in the $y$ direction. Thus, the system has attractive long-range interactions in the $y$ direction and repulsive ones in the $x$ direction (Fig. \ref{fig:ddi}). To investigate the system, we use the unbiased quantum Monte Carlo method based on the worm (directed-loop) algorithm\cite{prokofev1998, syljuasen2002, kato2009}. In our simulations, we treat systems of the size $N = L \times L$ with the periodic boundary condition. The lattice spacing is set to unity. To eliminate the effect of cutoff in long-range interactions, we employ the Ewald summation method\cite{ewald1921, deLeeuw1980}. In addition, we also adopt the $O(N)$ method\cite{fukui2009} to perform efficient simulations of systems with long-range (but integrable) interactions.
	
	The paper is organized as follows. Section \ref{sec2} presents the ground-state phase diagram in the grand-canonical ensemble. The phase diagram includes a superfluid phase and a striped solid phase at half-filling as two main phases. In addition to these phases, we find regions where multi plateaus appear at commensurate fillings in the particle density for small hopping amplitudes. In Sec. \ref{sec3}, we study finite-temperature transitions in the superfluid state and the striped solid state at half-filling. By performing finite-size scaling analysis, we reveal their universality classes and critical temperatures. Sec. \ref{sec4} describes the region where multi plateaus are observed in the particle density. Finally, in Sec. \ref{sec5}, we summarize our results.

\section{\label{sec2}Ground-State Phase Diagram}
\begin{figure}[h]
		\includegraphics[width=9cm]{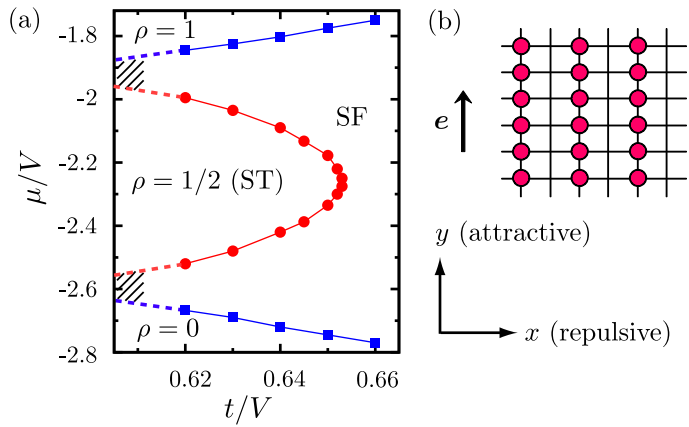}
		\caption{\label{fig:phasediag} (Color online) (a) Ground-state phase diagram of hard-core bosons on a square lattice with the fully anisotropic dipole-dipole interaction. In the present case, dipole moments are polarized in the $y$ direction. Error bars are drawn but most of them are smaller than the symbol size (here and the following figures). Shaded regions represent regions where we observe multi plateaus in the particle density for finite systems. Dashed lines are schematic phase boundaries. (b) Schematic configuration of the stripe solid state at half-filling.  Bosons are represented by circles.}
	\end{figure}

\begin{figure}[h]
		\includegraphics[width=7cm]{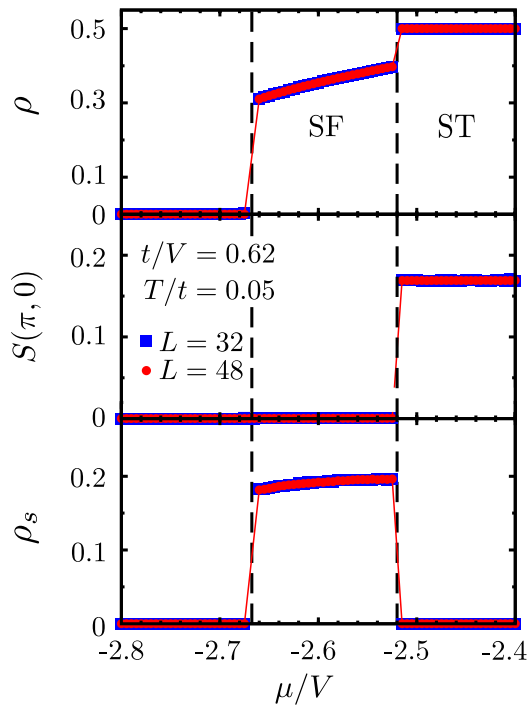}
		\caption{\label{fig:quantities} (Color online)  Particle density $\rho$, structure factor $S(\pi, 0)$, and superfluid stiffness $\rho_s$ as a function of the chemical potential $\mu/V$ at $(t/V, T/t)=(0.62, 0.05)$. Dashed lines separate the whole region into three different phases, namely empty phase, SF phase, and ST phase.}
	\end{figure}

	Our main result is the ground-state phase diagram shown in Fig. \ref{fig:phasediag}(a). In the phase diagram for $t/V \agt 0.62$, we find two phases, namely a superfluid (SF) phase and a striped solid (ST) phase at half-filling. The schematic configuration of the striped solid state is presented in Fig. \ref{fig:phasediag}(b). To detect each phase, we measure the particle density $\rho = 1/N \langle \sum_{i} n_{i} \rangle$, the superfluid stiffness $\rho_{s} = \langle \mbox{\boldmath $W$}^2 \rangle T/4t$, and the structure factor $S( \mbox{\boldmath $k$} ) = 1/N^2 \sum_{i, j} e^{i \mbox{\boldmath $k$} \cdot \mbox{\boldmath $r$}_{ij}} (\langle n_{i} n_{j} \rangle - \langle n_{i} \rangle^2 )$ at a low temperature $T/t=0.05$. Here, $\langle \cdots \rangle$ indicates the thermal expectation value, $\mbox{\boldmath $W$} = (W_{x}, W_{y})$ is the winding number vector in the world-line representation\cite{pollock1987}, and $\mbox{\boldmath $k$}$ is the wave vector.  In the striped solid phase, the ordering wave vector is $\mbox{\boldmath $k$}=(\pi, 0)$. In Fig. \ref{fig:quantities}, we plot the physical quantities as a function of the chemical potential $\mu/V$ at $(t/V, T/t)=(0.62, 0.05)$. The physical quantities show clear jumps at boundaries between two different phases, suggesting first-order transitions. In 2D systems with isotropic repulsive dipolar interactions, there is a theoretical prediction that first-order transitions with a density change are forbidden due to the negative log-divergent surface tension between two phases\cite{spivak2004}. In contrast, when dipole moments are pointing in the 2D plane, the surface energy becomes non-negative, and, therefore, first-order transitions are allowed\cite{knap2012}. For smaller hopping amplitudes $t/V \alt 0.61$, we have found regions where multi plateaus appear in the particle density for finite systems. The regions are depicted as shaded regions in Fig. \ref{fig:phasediag}(a). We present the numerical results on these regions and discuss the results in Sec. \ref{sec4}. In our simulation, we have found no evidence of a striped supersolid phase.
	
\begin{figure}[h]
		\includegraphics[width=8cm]{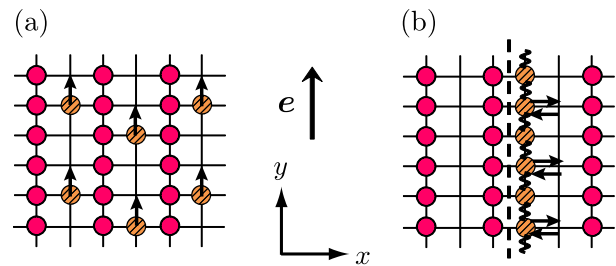}
		\caption{\label{fig:domainwall} (Color online) (a) Possible interstitial-induced supersolid state that is realized by delocalization of doped particles (shaded circles) on the striped solid background (simple circles). (b) A domain-wall (dashed line) formed by doped particles. Wavy lines represent the attractive nearest-neighbor interactions between doped particles. Arrows of doped particles indicate hopping process.}
	\end{figure}

	For small hopping amplitudes, the absence of a striped supersolid phase can be understood qualitatively by discussing its stability against domain-wall formations\cite{sengupta2005}. Although we consider the possibility of an interstitial-induced supersolid state in the following discussion, the same argument can also be applied to a vacancy-induced supersolid state because of the particle-hole symmetry. In Figs. \ref{fig:domainwall}(a) and (b), we present sketches of an interstitial-induced supersolid and domain-wall formation, respectively. We assume that the supersolid state is realized by Bose-Einstein condensation of interstitials on top of the striped solid\cite{andreev1969, chester1970}. In both situations, particles with small density $\rho \sim 1/L$ are doped into the striped solid state at half-filling. We first consider the case of the classical limit $t=0$. When we focus on interactions between doped particles, we notice that the energetic cost of the domain-wall formation in Fig. \ref{fig:domainwall}(b) is lower than that of the supersolid state in Fig. \ref{fig:domainwall}(a). This is because doped particles lower the energy by aligning in the attractive direction. In contrast, the interstitials in Fig. \ref{fig:domainwall}(a) are interacting weakly, because they are far from each other. Even if we consider the effect of sufficiently small hopping amplitudes, we expect that doped particles still prefer the domain-wall formation because of the large energetic gain in the zero-th order of $t$. Therefore, with doping of infinitesimal particle density, a supersolid state is expected to be unstable against the domain-wall formation for sufficiently small hopping parameters. When the hopping amplitude is increased, the situation is more complicated. This is because, when interstitials delocalize [Fig. \ref{fig:domainwall}(a)], the kinetic energy gain is $O(t)$, while it is only $O(t^2)$ in the case of the domain-wall formation [Fig. \ref{fig:domainwall}(b)]. This causes a possibility that, for finite hopping amplitudes, the kinetic energy gain overcomes the loss in the zero-th order of $t$, and, thus, the supersolid state becomes stable against the domain-wall formation. However, the absence of supersolid phase in our numerical results denies such a scenario in the present case.

	\section{\label{sec3}Finite-Temperature Transitions}
		In this section, we study finite-temperature transitions in the two main phases, namely ST phase and SF phase. By performing finite-size scaling analysis, we clarify their universality classes and critical temperatures. The results for ST phase and SF phase are presented in Sec. \ref{subsec3-1} and Sec. \ref{subsec3-2}, respectively.
		\subsection{\label{subsec3-1}Striped solid phase at half-filling}
\begin{figure}[h]
		\includegraphics[width=9cm]{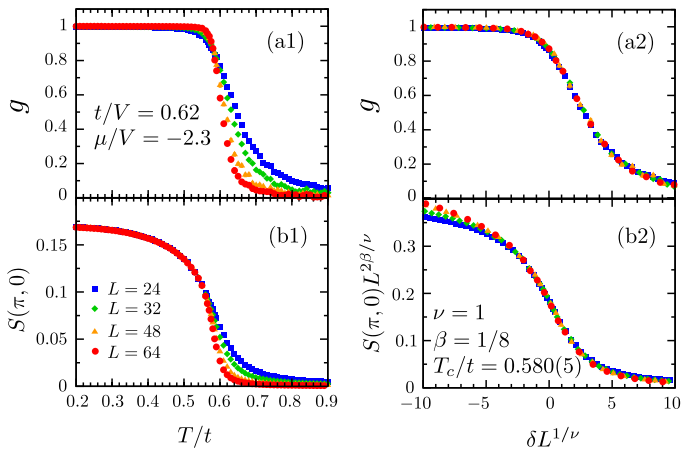}
		\caption{\label{fig:fss_stripe} (Color online) (a1) and (b1) Binder ratio $g$ and structure factor $S(\pi, 0)$ as a function of the temperature $T/t$ at $(t/V, \mu/V)=(0.62, -2.3)$, respectively. (a2) and (b2) Finite-size scalings of $g$ and $S(\pi, 0)$, respectively.}
	\end{figure}

			In this subsection, we study the finite-temperature transition to the ST phase. To this aim, we measure the Binder ratio $g=1/2[3-\langle m^4 \rangle/\langle m^2 \rangle^2$] as well as the structure factor $S(\pi,0)$. Here, $m$ is the order parameter defined by $m=1/N \sum_{i} n_{i} e^{i (\pi, 0) \cdot \mbox{\boldmath $r$}_{i}}$. We plot the Binder ratio $g$ and the structure factor $S(\pi,0)$ as a function of the temperature $T/t$ at $(t/V, \mu/V)=(0.62, -2.3)$ in Figs. \ref{fig:fss_stripe}(a1) and (b1), respectively. Both quantities take finite values at low temperatures in the ST phase. The critical temperature $T_{c}$ is estimated as $T_{c}/t=0.580(5)$ from the crossing point of curves of $g$ for different system sizes. To clarify its universality class, we perform the finite-size scaling analysis by using the scaling forms of $g=f(\delta L^{1/\nu})$ and $S(\mbox{\boldmath $k$}) L^{2\beta/\nu} = h(\delta L^{1/\nu})$. Here, $f$ and $h$ are scaling functions, and $\delta$ is defined by $\delta = (T-T_{c})/T_{c}$. $\nu$ and $\beta$ are the critical exponents. Since the phase transition is related to translational $Z_{2}$ symmetry breaking in the repulsive direction, the Ising-type universality class is expected. Therefore, in our scaling analysis, we use the critical exponents $\nu=1$ and $\beta=1/8$ (the 2D Ising universality class) as well as the obtained critical temperature. Figures \ref{fig:fss_stripe}(a2) and \ref{fig:fss_stripe}(b2) show the results of our finite-size scaling analysis for $g$ and $S(\pi,0)$, respectively. We successfully observe good data collapses that strongly supports our expectation.

		\subsection{\label{subsec3-2}Superfluid phase}
		
\begin{figure}[h]
		\includegraphics[width=7cm]{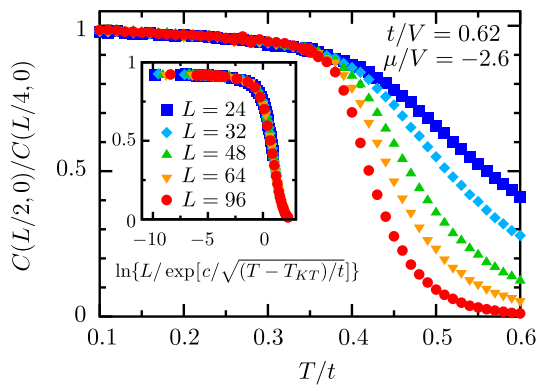}
		\caption{\label{fig:correlratio} (Color online) Correlation ratio $C(L/2,0)/C(L/4,0)$ as a function of the temperature $T/t$ at $(t/V, \mu/V)=(0.62, -2.6)$. In the inset, the result of finite-size scaling plots is shown.}
	\end{figure}

			We next study the finite-temperature transitions to the SF phase. The measured quantity is the correlation ratio $C(L/2,0)/C(L/4,0)$ in the $x$ direction. Here, $C(\mbox{\boldmath $r$})$ is the off-diagonal (superfluid) correlation function defined by $C(\mbox{\boldmath $r$})=\langle b_{\mbox{\boldmath $r$}} b^{\dagger}_{0} \rangle$. Figure \ref{fig:correlratio} shows the correlation ratio $C(L/2,0)/C(L/4,0)$ as a function of the temperature $T/t$ at $(t/V, \mu/V)=(0.62, -2.6)$. We observe the merge of the data in the SF phase, which is characteristic of the Kosterlitz-Thouless(KT)-type superfluid\cite{kosterlitz1973, ohgoe2011}. To estimate the critical temperature, we have performed the finite-size scaling analysis for the KT transitions. The scaling form is assumed to be $C(L/2,0)/C(L/4,0)=f(L/\exp[c/\sqrt{(T-T_{\rm KT})/t}])$\cite{kosterlitz1974, tomita2002}. Here, $c$ and the critical temperature $T_{\rm KT}$ are free parameters in the present analysis. The result is shown in the inset of Fig. \ref{fig:correlratio}. In the analysis, we have estimated the unknown values as $c=1.17(27)$ and $T_{\rm KT}/t=0.334(11)$. We have also performed the similar analysis for the correlation ratio $C(0,L/2)/C(0,L/4)$ in the $y$ direction, and obtained a consistent critical temperature within the error bar (not shown here). 

	\section{\label{sec4}Multi Plateaus in the Particle Density}

\begin{figure}[h]
		\includegraphics[width=6.5cm]{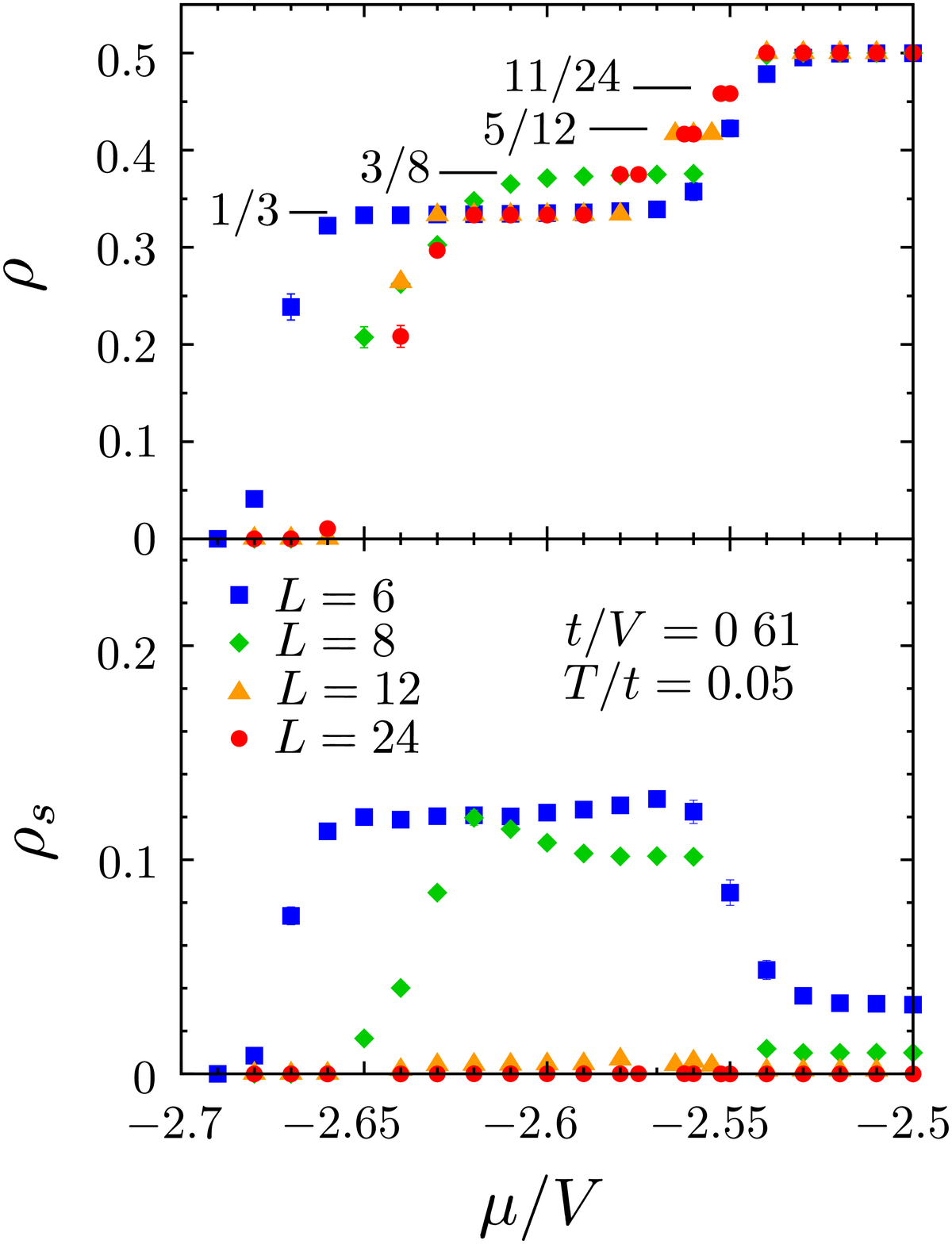}
		\caption{\label{fig:staircase} (Color online) Particle density $\rho$ and superfluid stiffness $\rho_{s}$ as a function of the chemical potential $\mu/V$ at $(t/V, T/t)=(0.61, 0.05)$.}
	\end{figure}

			In this section, we present numerical results for shaded regions shown in Fig. \ref{fig:phasediag}(a). In these regions, behaviors of the physical quantities are neither those of SF phase nor ST phase. To show it, we plot the particle density $\rho$ and the superfluid stiffness $\rho_{s}$ as a function of the chemical potential $\mu/V$ at a low temperature $T/t=0.05$ in Fig. \ref{fig:staircase}. The particle density $\rho$ shows multi plateaus at commensurate values and the number of plateaus increases as the system size increases. On the other hand, the superfluid stiffness $\rho_{s}$ is suppressed there and the value decreases rapidly as the system size increases. In the narrow regions between two adjacent plateaus ($e.g.$ $\rho=1/3$ and 3/8 for $L=24$), we have found that the system reaches different adjacent commensurate states according to initial states. This suggest that a direct transition occurs between two adjacent phases with commensurate fillings and there is little possibility of any phase with superfluidity. Unfortunately, we have found for larger system size than $L=24$ that it is difficult to detect more plateaus clearly due to the presence of numerous metastable states\cite{menotti2007}.

\begin{figure}[h]
		\includegraphics[width=9cm]{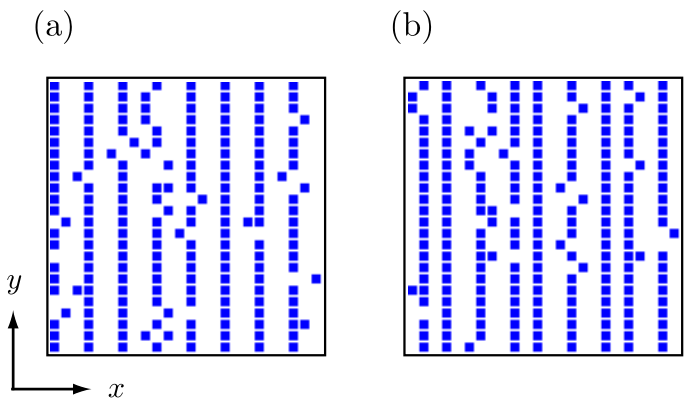}
		\caption{\label{fig:snapshot} (Color online) (a) and (b) Snapshots of particle configurations in a real space at plateaus $\rho=1/3$ and $\rho=3/8$, respectively. The parameters are chosen at $(L, t/V, \mu/V, T/t)$ = (24, 0.61, -2.6, 0.05) and (24, 0.61, -2.575, 0.05), respectively. Each site is denoted as a square. Open and blue squares indicate empty and occupied sites, respectively. Both snapshots show almost striped structures with defects that derive from quantum and thermal fluctuations.}
	\end{figure}

\begin{figure}[h]
		\includegraphics[width=6cm]{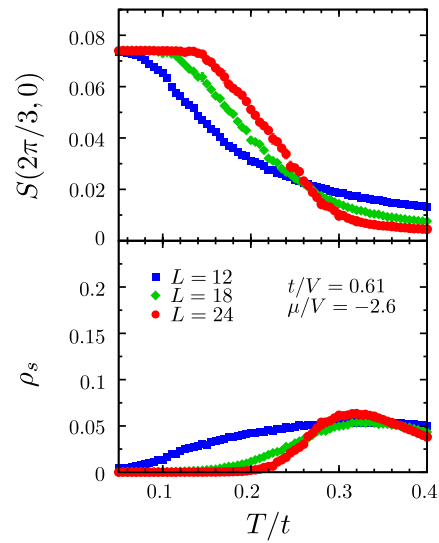}
		\caption{\label{fig:one-third_stripe} (Color online) Structure factor $S(2\pi/3, 0)$ and superfluid stiffness $\rho_{s}$ as a function of the temperature $T/t$ at $(t/V, \mu/V)=(0.61, -2.6)$.}
	\end{figure}

			To reveal configurations of the commensurate states, we present snapshots of the particle configuration. Figures \ref{fig:snapshot}(a) and (b) show the results at plateaus $\rho=1/3$ and 3/8, respectively. The state at $\rho=1/3$ has a striped structure with periodicity of 3 in the $x$ direction. To confirm it quantitatively, we have calculated $S(2\pi/3, 0)$ and $\rho_{s}$. These quantities are plotted as a function of the temperature in Fig. \ref{fig:one-third_stripe}. In the systems of $L=$12, 18, and 24, the values of $S(2\pi/3, 0)$ are independent of the system sizes at low temperatures, whereas the value of superfluid stiffness $\rho_{s}$ decreases rapidly as the system size increases. (At higher temperatures $T/t \sim 0.3$, a SF phase may exist, because the value of superfluid stiffness $\rho_{s}$ does not vanish within the present system sizes.) This indicates that a striped solid state with periodicity of 3 is a possible ground state. However, it may change to the other commensurate state in larger systems at $\mu/V=-2.6$, because the width of plateau becomes smaller as the system size increases. The striped solid order is mainly caused by the next-nearest-neighbor repulsion in the $x$ direction. Similarly, the snapshot of the state at $\rho=3/8$ also shows almost striped structure and can be well explained by mixture of stripes with periodicity of 2 or 3 in the $x$ direction. The appearance of the striped state at $\rho=3/8$ is due to the competition of the nearest-neighbor and next-nearest-neighbor repulsions in the $x$ direction. This competition implies emergence of numerous striped states with different combinations of the periodicity 2 and 3 in large systems, and, thus, gives rise to corresponding plateaus in the particle density. From the present data, it is impossible to determine whether, in the thermodynamic limit, it results in an infinite sequence number of commensurate phases (devilfs staircase) or incommensurate phases\cite{selke1979, fisher1980, bak1982, burnell2009, capogrosso2010-1} because of the strong system size dependence. For the same reason, the precise phase boundaries remain unclear. Therefore, we simply denote it as the shaded region in the ground-state phase diagram [Fig. \ref{fig:phasediag}(a)].

\section{\label{sec5}Summary}
	We have investigated the hard-core Bose-Hubbard model on a square lattice with fully anisotropic dipole-dipole interaction by using unbiased quantum Monte Carlo method. The ground-state phase diagram includes a superfluid phase and a striped solid phase at half-filling as two main phases. Furthermore, for small hopping amplitudes, we observe small regions where there are multi plateaus at commensurate fillings for finite systems. In the present case, a striped supersolid phase does not appear because of the strong attractive interactions in the $y$ direction. Such a striped supersolid phase might be observed in other cases such as negative $V$\cite{yi2007} or a different direction of dipole moments \cite{danshita2009}. 
	
\section*{ACKNOWLEDGMENTS}
	The authors are grateful to D. Yamamoto, I .Danshita, and Y. Tomita for valuable discussions. The present work is financially supported by the Global COE Program ``the Physical Science Frontier", a Grant-in-Aid for JSPS Fellows (Grant No. 249904), a Grant-in-Aid for Scientific Research (B) (22340111), a Grant-in-Aid for Scientific Research on Priority Areas ``Novel States of Matter Induced by Frustration" (19052004), and the Computational Materials Science Initiative (CMSI), Japan. The simulations were performed on computers at the Supercomputer Center, Institute for Solid State Physics, University of Tokyo.


\end{document}